\documentstyle[aps,prl,multicol,epsfig]{revtex} 
\begin{document}
\draft
\title{ Berry phase theory of the Anomalous Hall Effect:
  Application to Colossal Magnetoresistance Manganites}
\author{Jinwu Ye$^1$, Yong Baek Kim$^2$, A. J. Millis$^3$, B. I. Shraiman$^4$, 
 P. Majumdar$^5$ and Z. Te\v{s}anovi\'{c}$^1$ }
\address{ $^1$Department of Physics and Astronomy,
The Johns Hopkins University, Baltimore MD, 21218  \\
  $^2$Department of Physics, The Pennsylvania State University,
   University Park, PA 16802  \\
  $^3$Department of Physics, Rutgers University, Piscataway, NJ 08854  \\
  $^4$Bell Laboratories, Lucent Technologies, Murray Hill, NJ 07974\\
  $^5$Mehta Research Institute, Chatnag Road, Jhus, Allahabad, India 221-506 }
\date{\today}
\maketitle
\begin{abstract}
  
  We show that the Anomalous Hall Effect (AHE) observed in
  Colossal Magnetoresistance Manganites is a manifestation of Berry phase effects caused by carrier hopping in a non-trivial spin background.
  We determine the magnitude and temperature dependence of the Berry
  phase contribution to the AHE,
  finding that it increases rapidly in magnitude as the temperature is raised
  from zero through the magnetic transition temperature $ T_{c} $, peaks
  at a temperature $ T_{max} > T_{c} $ and decays as a power of T,
  in agreement with experimental data. We suggest that our theory
  may be relevant to the anomalous Hall effect
  in conventional ferromagnets as well.

\end{abstract}

\begin{multicols}{2}

    The Anomalous Hall Effect(AHE) is a fundamental but incompletely understood
    aspect of the physics of metallic ferromagnets \cite{chien,hurd}.
    The Hall effect is the development of a voltage which is transverse
    to an applied current; the constant of proportionality is the Hall
    resistivity $ \rho_{H} $. In non-magnetic materials, $ \rho_{H} $
    is proportional to the magnetic induction $ B $ and its sign is 
    determined by the carrier charge. Many ferromagnets however exhibit
    an anomalous contribution to $ \rho_{H} $ which is proportional to
    the magnetization $ M $, thus
\begin{equation}
   \rho_{H}= R_{0} B + R_{s} M
\label{def}
\end{equation}

    The definition of $ R_{s} $ implies a sample with
    demagnetization factor $ N \cong 1 $ so that $ M $ represents the spin
    polarization in the material and the physical dipolar magnetic field
    caused by the ferromagnetically aligned spins cancels. 
    The AHE thus involves a 
    coupling of orbital motion of electrons to the spin polarization
    and must involve spin-orbit coupling.

    The conventional theoretical understanding of $ R_{s} $ is based 
    on a skew scattering 
     mechanism which is a third order process involving interference between 
     spin-orbit coupling (to first order ) and spin flip scattering ( to
     second order ) \cite{chien,so,semi}. In conventional ferromagnets, this
     theory yields values of $ R_{s} $
     two orders of magnitude smaller than experimental data \cite{so}.
     Also, some papers including
     Ref.\cite{so} use a spin-orbit term involving coupling to the 
     dipole fields
     produced by the spins which  would apparently vanish for
     demagnetization factor $ N=1 $.
     
      Recently, several groups measured the
    Hall resistivity $ \rho_{H} $ of epitaxial films \cite{ong} 
    and single   crystals \cite{chun} of  the 'colossal magnetoresistance'
    (CMR) material $ La_{0.7} Ca_{0.3} Mn O_{3} $ 
    These materials involve carriers derived from Mn $ e_{g} $
	symmetry d-levels which may move through the lattice but
	are strongly ferromagnetically coupled to localized 'core spins'
	derived from Mn $t_{2g} $ symmetry orbitals. The coupling is so 
	strong that it may be taken to be infinite: a carrier on site i
	must have its spin parallel to the core spin on site i. The spin
	of the mobile carrier is thus quenched, but its amplitude to hop
	from site i to site j is modulated by a factor involving the relative
	spin states of core spins on the two sites. This physics is called
	'double-exchange'\cite{anderson}.
	
	The Hall effect measurements found 
	that in CMR materials $ \rho_{H} $ was of the form of
    Eq.\ref{def} with $ R_{0} $ {\em hole-like} and $ R_{s}  $ 
    {\em electron-like}.
    $ R_{s} $ becomes evident above 100K,  increases sharply 
    around $ T_{c}  $, peaks at a temperature  $ T_{max} \approx T_c+30K$ 
  and decreases slowly at high temperatures. d
    $ R_{s} $ is proportional
    to the zero-field resistivity from 200K to 360K.
    This cannot be explained via
    the conventional skew scattering theory because the quenching of
    the carrier spin means the required spin-flip process cannot
    occur.  That  the sign of $ R_{s} $ is {\em opposite } to $ R_{0}
    $, and that
    $ R_{s} $ peaks {\em above} $ T_{c} $ are also surprising.

       In this paper, we present a new theory for
       anomalous Hall effect. It is inspired by the physics of CMR, 
       but we suggest that a simple generalization could apply to 
       conventional ferromagnets as well. Our mechanism is based on the
       observation that a carrier moving in a topologically non-trivial
       spin background acquires a 'Berry phase'\cite{Schulz90} which
       affects the motion of electrons in the same way as does the phase 
       arising from a physical magnetic field \cite{Ioffe89} and has
       been argued  to influence the Hall effect in high-$T_c$
       superconductors \cite{Ioffe91}. We shall
       show that this Berry phase can in the presence of spin-orbit
       coupling give rise to an AHE similar
       in magnitude and temperature dependence to that observed in 
       CMR materials. This idea was advanced in a preprint \cite{kim}. The 
        present paper treats skyrmion
	physics and the high temperature phase in more detail, and 	
        presents a more general treatment valid for any half-metallic
	ferromagnet.  It supercedes the previous work.


After the work presented here was completed two preprints appeared  
presenting  a discussion of the physics introduced in \cite{kim} along
with  new data which are argued to confirm the basic 
picture \cite{lyanda99}.

    We begin our analysis by writing down the DE model for electrons 
hopping on a lattice, coupled ferromagnetically by atomic exchange
$ J_{H} $ to core spins $ S_{c} $ and subject to a spin-orbit
coupling $ H_{so} $ and to other interactions represented by
 $ \cdots $:
\begin{eqnarray}
  H &=& -t \sum_{<ij> \alpha}
  ( d^{\dagger}_{i  \alpha} d_{j  \alpha} + h. c. )
      - \mu \sum_{<ij> \alpha}  d^{\dagger}_{i  \alpha} d_{j  \alpha} \cr 
    &&-J_{H} \sum_{i} \vec{S}_{ic} \cdot d^{\dagger}_{i  \alpha}
      \vec{\sigma}_{\alpha \beta} d_{i  \beta} + H_{so}+ \cdots
\label{de}
\end{eqnarray}
    The orbital indices of two $ e_{g} $ orbitals are suppressed in
    Eq. \ref{de} because they are not essential in this paper
    \cite{orbital}.
    $ H_{so} $ in Eq.\ref{de} arises from the fundamental spin-orbit coupling
   $ H_{so} \sim (\vec{k} \times \vec{\sigma} ) 
   \cdot  \vec{\nabla} V_{c} $ ($ V_{c} $ is the crystalline potential
   and $ \vec{\sigma} $ are the Pauli spin matrices). The projection
    of $H_{so}$ onto tight binding bands leads to many terms
    \cite{semi}; the one of relevance here is
\begin{equation}
  H_{so}=i\frac{\lambda_{so}}{4} \sum_{i}
  \epsilon^{abc} \sigma^{a}_{\alpha \beta}
  ( d^{\dagger}_{i+b, \alpha} - d^{\dagger}_{i-b, \alpha} )
  (d_{i+c,\beta}- d_{i-c, \beta} )
\label{orig}
\end{equation}
  
     The crucial physics of the manganites is a strong Hunds coupling
     $ J_{H} \gg t $.
    In the following, we study the $ J_{H}/t \rightarrow \infty $
     limit, and comment on the $J_H <t$ case in the conclusion.
    It is convenient to parameterize $ \vec{S}_{ic} $ by polar 
    angles $ \theta_{i}, \phi_{i} $ and to express the electron operator as 
\begin{equation}
   d_{i  \alpha}= d_{i } z_{i  \alpha},~~~~~ 
  z_{i \alpha}= |\vec{n}_{i}> = \left( \begin{array}{c}
   \cos \frac{\theta_{i}}{2} \\  \sin \frac{ \theta_{i} }{2} e^{i \phi_{i}}
   \end{array} \right)
\label{sep}
\end{equation}
   where $ z_{i \alpha} $ is the $ SU(2) $ coherent spin state
   along $ \vec{n}_{i}= z^{\dagger}_{i \alpha}
	\vec{\sigma}_{\alpha \beta} z_{i \beta} $.

  Using Eq.\ref{sep}, in the presence of an external magnetic field
  $ \vec{H}=\nabla \times \vec{A} $, 
  we can write the  action 
\begin{eqnarray}
  {\cal S} &=&  \int^{\beta}_{0} 
   d \tau \{ \sum_{i} i n_{c} a_{0}
   + \sum_{i}   d^{\dagger}_{i} ( \partial_{\tau} - i a_{0} - \mu) d_{i} \cr
    &-&  t \sum_{i \hat{\delta}} ( \frac{1+ \vec{n}_{i}
   \cdot \vec{n}_{j} }{2} )^{1/2}
    [ e^{i a (a_{i \hat{\delta} }+ \frac{e}{\hbar c}
    A_{i \hat{\delta} })} d^{\dagger}_{i}
    d_{i+\hat{\delta}} +h.c. ] \cr
    &-& g \mu_{B} \vec{H} \cdot \sum_{i} \vec{n}_{i}
    ( S_{ic} + \frac{1}{2} d^{\dagger}_{i} d_{i} ) \} + H_{so}
\label{lattice}
\end{eqnarray}
    where $ n_{c}=2 S_{c} $, $ \mu $ is the chemical potential
    fixing the electron density $ < d^{\dagger}_{i} d_{i}>=x $.
    $ a $ is the lattice constant,
    $ a_{0}= i z^{\dagger} \partial_{\tau} z$ and $a_{i,\delta}$
    (defined below) 
    are the internal gauge fields generated by the  spin
    configurations.
      
      The term involving $ t $ shows explicitly how the electron hopping
      is affected by the nearest neighbor spin overlap factor
       $ z^{\dagger}_{i \alpha} z_{j \alpha} 
      =< \vec{n}_{i}|\vec{n}_{j}>
    =e^{i \Phi(\vec{n}_{i},\vec{n}_{j})/2 }( \frac{1+ \vec{n}_{i}
    \cdot \vec{n}_{j} }{2} )^{1/2}  $. The phase factor 
    $  \Phi(\vec{n}_{i},\vec{n}_{j})= a \ a_{i \hat{\delta}} \
    (j=i+\hat{\delta}) $ is the solid angle subtended
   by the three unit vectors $ \vec{n}_{i}, \vec{n}_{j} $ and
   $ \hat{z} $ on the unit sphere. In the continuum $a_{i,\delta}
      \rightarrow z^{\dagger} \partial_{\delta} z$.  $a_{i,\delta}$ 
    affects the motion of a electron
   just as does a external electromagnetic field \cite{Ioffe89}. 
   A time dependent $\vec{a}$ was shown 
   by Nozieres and Lewiner \cite{semi} to lead to a time dependent AHE 
       when the magnetization
       precessed.  Here we show  leads to a static AHE when
      topologically nontrivial spin configurations are considered.

      We now use Eq. \ref{lattice} to
        calculate  $ R_{s} $, considering the  low, high 
       and critical temperature regimes separately.

\underline{ The low temperature regime:} here the core spins are slowly
   varying on the lattice scale and the system is metallic,
   so we can take the continuum limit, treat the core spins classically 
   and obtain
\begin{eqnarray}
   {\cal S} &=& \beta \int d^{d}x [ \frac{\rho_{s}}{2}
   (\partial_{i} \vec{n})^{2} 
   - M_{0} \vec{H} \cdot \vec{n} ] + S_{el}+H_{so}
\label{mag}
\end{eqnarray}
    where $ \rho_{s} \sim tx a^{2-d} $ is the spin stiffness,
    $ M_{0}= g \mu_{B} [ S_{c} +\frac{1}{2}x ] a^{-d} $ is the magnetization,
     and
    $ S_{el} $ is the action of spinless fermions moving 
    in the band structure defined by $ t $ in Eq.\ref{de} and
    coupled to a gauge field
    $ \vec{a} + \frac{2 \pi}{ \Phi_{0} } \vec{A} $, to spin
    fluctuations  and 
    to other interactions not explicitly written.
    It is important to stress that in the CMR case considered here,
    the core spins and
    electrons are independent fields and the internal
    gauge field is simply a representation of topologically non-trivial 
    configuration of the core spins instead of an {\em independent} field. 
    However, in high $ T_{c} $ cuprates,
    the gauge fields are {\em independent } fields related to
    spin-charge separation \cite{Ioffe89}.

Taking the continuum limit of Eq.\ref{orig} and neglecting terms involving
electron currents yields

\begin{equation}
  H_{so}= i \lambda_{so}xa^2 \epsilon^{abc} 
  (\partial_{b}+i a_{b}) z^{\dagger}_{\alpha}\sigma^a_{\alpha,\beta} 
  (\partial_{c}-i a_{c}) z_{\beta}
  \label{Hsocontinuum}
\end{equation}
 Integrating the conduction electrons out of Eq.\ref{mag}, and 
 evaluating $H_{so}$ using Eqs.\ref{sep},\ref{Hsocontinuum} yields
\begin{eqnarray}
   {\cal S}  &=&  \beta \hbar \int d^{d}x [ \frac{\rho_{s} }{2}
   (\partial_{i} \vec{n})^{2} - M_{0} \vec{H} \cdot \vec{n}+
   \frac{\chi_{F}}{2} ( \frac{2 \pi}{ \Phi_{0} } \vec{H} +  \vec{b} )^{2} ] \cr
  &&+ \lambda_{so} \frac{x}{2a}
       \int d^{d}x  \vec{n} \cdot \vec{b}
\label{split} 
\end{eqnarray}
    where $ \chi_{F} \sim ta x^{1/3} $ 
    is the electron diamagnetic susceptibility,
     $\Phi_{0} = \frac{hc}{e} =4.1358 \times 10^{-7}
    gauss \cdot cm^{2} $ is the fundamental flux quantum and
    $ b_{i}= \epsilon_{ijk} 
    \partial_{j} a_{k}= \frac{1}{4} \epsilon_{ijk}
    \vec{n} \cdot (\partial_{j} \vec{n} \times \partial_{k} \vec{n} ) $
    is the internal magnetic field arising from the gauge field $\vec{a}$.
        
 A non-zero $ b $ arises from topologically non-trivial spin configurations.
 In two spatial dimensions, these are the skyrmions \cite{belavin}
 which are important  in  quantum
 Hall ferromagnets \cite{sky}.
    In three dimensionals, the objects are skyrmion strings (dipoles)
    which begin at monopoles and end at an anti-monopoles.
    In the ordered phase, these  have finite
    creation energies and  are  exponentially suppressed
    at low T, but can proliferate near $ T_{c} $. Their behavior
    has been studied numerically \cite{lau}.

     It is known that at zero external magnetic field,
     the core energy of a dipole separated by distance
     $ d $ is $ E_{c}=4 \pi \rho_{s} d $ (i.e. string tension 
      $\sim \rho_s$).
    To relate $E_c$ to experimental paramters we note that according to
    the numerical analysis of Ref.\cite{lau}, 
    $ T_{c} = 1.45 \rho_{s} a $, while in the experiment of
     Ref. \cite{ong} $ T_{c}=265K $.
    We find the core energy for a dipole separated by a lattice constant
    $ a $ is $ E_{c} =2295 K $.
     Following Ref.\cite{sky}, we find only $ 2-3 \% $ increase of  $ E_{c} $
     in an external magnetic field $ H=10 T $, therefore, we can
     simply neglect the core energy dependence of $ H $.
    The core energy $ E_{c} $ 
    is {\em independent}
    of the angle $ \phi $ which defines the global orientation of the XY spin
    component. This $U(1)$ invariance implies the existence of a
     family of very soft twist modes.  Dilatation modes have a small energy
     gap but are also likely to be thermally excited. 
     These soft modes mean that at $T>0$ each skyrmion carries a
     large entropy.

    In the absence of spin-orbit coupling, dipoles are randomly
oriented, leading to vanishing average b.  In the presence of
spin-orbit coupling and a non-zero magnetization, the dipoles are
preferentially oriented, leading to a nonzero average b. 
To see this we rewrite 
Eq.\ref{split} in terms of the polarization 
$\vec{P}=Q\vec{l}$ where $Q=\pm 1$ is the charge of a monopole and
   $ \vec{l} $ is the vector connecting
   the monopole to anti-monopole which are the two end points of the dipole. 
   We find, for $\vec{M}$ parallel to {\bf z}
\begin{equation}
   M_{\lambda}\int dz \int dx dy b_{z} = M_{\lambda}
   \int dz Q_{z}= \frac{\lambda_{so}x}{a}\frac{\vec{M}a^3}{g\mu_B} 
 \cdot \vec{P}
   \label{pol}
\end{equation}
   where $M_{\lambda}= \frac{\lambda_{so}x}{a} \frac{M a^3}{ g \mu_{B} }$ .
  
     At low $ T $, numerics  Ref.\cite{lau} shows that
     the monopole and anti-monopole are very dilute and are
     tightly bound into dipoles of size of one lattice constant. 
     We thus can treat them as independent
     classical particles. The density ($n_{\pm}$) of (anti-) skyrmions is
\begin{equation}
    n_{\pm}= \alpha e^{-\beta( E_{c} \pm \lambda M Qa)}
\end{equation}
     where $ \alpha $ is the entropy per dipole.
     Refs.\cite{lau} find
      $\alpha \sim 320$.  We believe this large factor 
      comes from the twist and
     dilatation modes  mentioned above.

      The average internal magnetic field $<b>=-Q(n_+-n_-)/a^2$ is thus
\begin{eqnarray}
    \left < b \right > 
  &=&- \alpha 
     \frac{Q}{ a^{2}} e^{- \frac{ E_{c} }{ k_{B} T} }
     \sinh( \frac{Q \lambda_{so}x Ma^3 }{ k_{B} Tg\mu_B } ) 
\label{diff}
\end{eqnarray}
    
     An internal field $b$ produces the same change in the phase of an
     electron as would be produced by a physical field of magnitude
     $\frac{\Phi_0}{2\pi}b$.  The equivalent physical field is large
     because the appropriate dimensionless coupling (fine structure
     constant)  for the internal
     field is 1, rather than the 1/137 relevant for physical fields.

     The internal magnetic field produces a Hall effect
     in the usual way.  Writing the Hall resistivity as 
     $ \rho_{H} = \frac{1}{ n_{eff} e c} ( H+ 
     \frac{\Phi_{0}}{2 \pi} \left < b \right > ) $, 
     and linearizing the $ \sinh $, we find
\begin{equation}
   \frac{R_{s}}{ R_{0}} =- \alpha
   \frac{\Phi_{0}}{2 \pi} \frac{ \lambda_{so} Q^{2}}{  k_{B} T }
   \frac{a x}{g \mu_{B}} e^{- \frac{ E_{c} }{ k_{B} T} }
\label{linear}
\end{equation}
      $ Q =1 $ is the charge of a monopole, $ a=3.92 \stackrel{\circ}{A} $ is
      the lattice constant of $ La_{1-x} Ca_{x} Mn O_{3} $, 
      At $ T=200 K $, the experimental value \cite{ong} is
      $ \frac{ R_{s}}{ R_{0} }= -13.6 $, implying 
      $ \frac{ \lambda_{so} }{ k_{B} T }= 2.5 \times 10^{-2} $ so
      $ \lambda_{so} \sim 5 K $ which
      justifies the linearization used in Eq.\ref{linear}.
      Because the on-site spin-orbit
      interaction is quenched by the cubic crystal field, $\lambda_{so}$
      is dominated by inter-ion hopping and 
      is difficult to determine, but a rough estimate can be
      obtained by combining the dimensionless coupling
      appropriate for d-orbitals ($\frac{Ze^2}{2mc^2a_0}$) ($a_0 \sim
      0.5 \AA$ is
      the d-orbital size) with the band kinetic energy
      $\frac{\hbar^2}{2ma^2}$ yielding $\lambda_{so} \sim 2K$.  
     
     The sign of $R_s$ relative to $R_0$ depends on the sign of the
     spin orbit coupling.  The physically reasonable
     sign (usual precession of charge around field of core spin) leads
     to an internal field $b$ which acts to cancel the applied field,
     implying opposite signs for $R_s$ and $R_o$.

\underline{ The high temperature regime:} here 
    $ k_BT_c, g \mu_{B} H, \ll k_{B} T \ll J_{H} $, so
    the core spins
    vary on the lattice scale,  the continuum
    action Eq.\ref{split} does {\em not} apply and the skyrmion is not
    well defined.  
    We calculate the average of the z-component of the gauge invariant
    internal flux  
    $ \Phi_{\triangle}=\epsilon^{abc}n^a_{i+\hat{x}}n^b_{i}n^c_{i+\hat{y}}$ 
   through a
   triangle of adjacent sites $i+\hat{x},i,i+\hat{y}$ to leading order
    in $\lambda_{so}/k_BT$ using the lattice action
   Eq.\ref{lattice}, finding 
\begin{equation}
	\Phi_{\triangle} \sim \frac{\lambda_{so}\chi H}{k_BT}
	\sum_{\alpha}\langle d^{\dagger}_{i+x,\alpha}d_{i+y,\alpha}\rangle
\label{phibox} 
\end{equation}
  where $\chi$ is the spin susceptibility.

   Further, the $La_{1-x}Ca_xMnO_3$ materials of main
    experimental interest are in a highly resistive 'polaron hopping'
    regime which we model  by assuming $t \ll k_bT$. In the original
    lattice model with only nearest neighbor hopping
    the fermion  expectation value is of second order in
    $t/k_BT$, and may be computed, leading to 
%
\begin{equation}
   \frac{R_{s}}{ R_{0}} = -\frac{\lambda_{so}}{k_BT}  ( \frac{t}{k_{B} T})^{2} 
     x(1-x)(1-2x)C_{3} 
  \label{tri}
\end{equation}
   where $ C_{3} $ is a constant of order unity. (Recall 
   $R_s$ is the term in the Hall resistance proportional to
   $M \sim \chi H$.)


  
  The difficulty of unambiguously extracting the term in 
  the Hall conductivity proportional to the
  spin susceptibility increases
  as $ T $ is increased further above 
  $ T_{c} $, so a detailed study of the high-$T$ $R_s$ is
  probably not waranted.

\underline{ The critical regime:} For $ T $ near $ T_{c} $,
       a theory of the topological defects  has not been
       constructed, because in the
      3d Heisenberg model, there is 
     {\em no} decoupling between spin-wave fluctuations and the topological
     defects, so spin waves cannot simply be integrated out.
    Ref.\cite{lau} simulated the behavior of the topological
    defects near $ T_{c} $ at zero external magnetic
    field. The dipole density $ n $ was found to increase sharply as $ T $
    passes through $ T_{c} $; there is a derivative discontinuity at 
   $T_c$ which is controlled by
    the specific heat exponent $\alpha$ \cite{lau}, 
   (as is also the case 
    for the resistivity). The singularity at $T_c$
may be difficult to observe:  it is cut 
off by the field needed to observe an AHE at $T>T_c$ and 
the AHE has also an amplitude involving the field 
and temperature dependence of the 
magnetization. 



    From the low temperature and critical regime calculations, we find
    $ R_{s} $ increases as $T$ is increased, and $dR_s/dT >0$ at $T_c$.
    The high temperature expansion yields a 
    $d R_{s}/dT <0 $. Therefore, 
    $ R_{s} $ has a maximum at some $T_{max}>T_c$.  The decrease for
    $T>T_{max}$  has two causes--strong thermal spin fluctuations
    disrupt  the local 
    correlations needed to produce a $b$, and also the fermion
    correlator in Eq \ref{phibox} decreases.  Our knowledge of CMR
    materials is insufficient to allow a quantitative theory of the 
    region near $T_{max}$.  

	To summarize:  we have constructed a new theory of the
    anomalous Hall effect in CMR manganites.  Our results
    are shown in the Figure, and account naturally
    for the order of magnitude, the sign (relative to the conventional
    Hall effect), and the peak at $T>T_c$ found experimentally    \cite{ong,chun}.

\vspace{0.25cm}

\epsfig{file=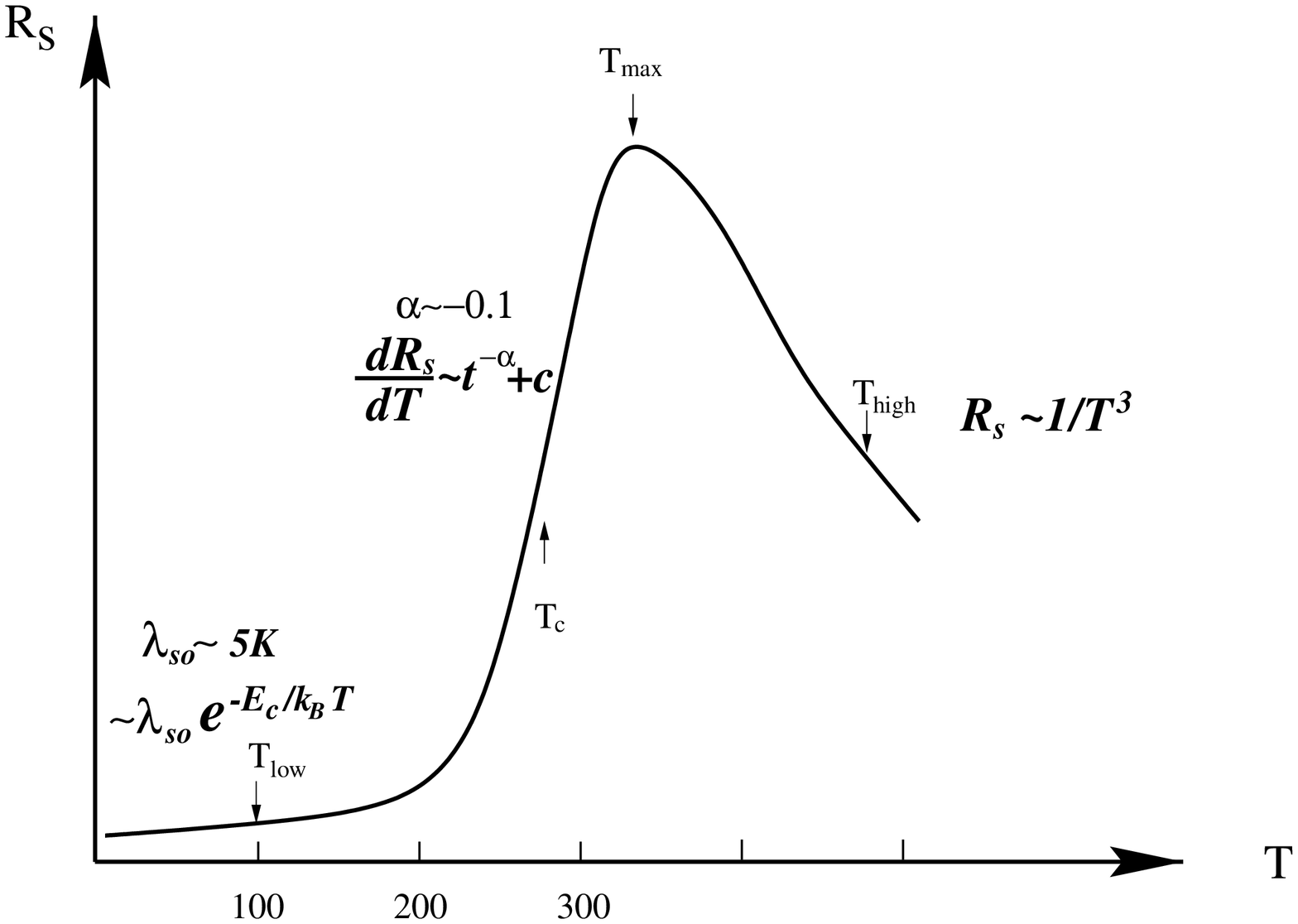,width=3.2in,height=1.9in,angle=0}

{\footnotesize {\bf Fig 1} Calculated temperature dependence of
    anomalous Hall coefficient}

\vspace{0.25cm}

       Our results for $R_s$ have same qualitative behavior as (and
    identical critical behavior to) the
    longitudinal resistivity, but we have no argument for the very close
    correspondence between the two found experimentally \cite{ong}.

    This paper has assumed a large $ J_{H} $ limit, so the conduction band
    is completely spin-polarized. An extension to the smaller $ J_{H} $ limit,
    with a partially-polarized conduction band 
    would be of interest in connection with the AHE in conventional
    magnets. In this case, conventional skew scattering would also
    contribute and indeed would dominate at very low $T$ because it
    varies as a power of $T$.     Recently, a very large 
    AHE was found in Co oxides \cite{yeh}.
      These materials involve additional physics, including
      high-spin/low-spin transitions and inhomogeneous states,
      whose incorporation in our theory would be of interest.

 We thank C. L. Chien, S. H. Chun, R. Fisch, S. Kivelson, G. Murthy, N. P. Ong,
 M. B. Salamon, A. M. Sengupta, S. H. Simon and N. C. Yeh
 for helpful discussions.
 This 
research was supported by NSF Grant No. DMR-97-07701 ( J. Ye \& A. M)
 A. P. Sloan Foundation Fellowship (Y. B. K) and DMR-9415549 ( Z. T).

\end{multicols}


\begin{references}
\bibitem{orbital} The quantum phase associated with the orbital degree
of freedom may also in some circumstances lead to an AHE; see
e.g. Y.B. Kim et. al., op. cit.
\end{references}
\end{document}